\documentclass[sn-standardnature,referee]{sn-jnl}

\jyear{2025}%
\theoremstyle{thmstyleone}%
\theoremstyle{thmstyletwo}%

\theoremstyle{thmstylethree}%
\usepackage{upgreek}

\usepackage{xcolor}
\usepackage{ulem}

\newcommand\uphbar{\mathord{\mathrm{\hbar}}}
\begin{document}
\title[Mechanical dissipation near STO]{Mechanical detection of sub-band mobilities of two-dimensional electron gas on reduced SrTiO$_3$(001) surface}


\author[1]{\fnm{Akash} \sur{Gupta}}
\equalcont{These authors contributed equally to this work.}

\author*[1]{\fnm{Marcin} \sur{Kisiel}}\email{marcin.kisiel@unibas.ch}
\equalcont{These authors contributed equally to this work.}

\author[1]{\fnm{Rémy} \sur{Pawlak}}

\author*[1]{\fnm{Ernst} \sur{Meyer}}\email{ernst.meyer@unibas.ch}

\affil[1]{\orgdiv{Department of Physics, WSS-Research Center for Molecular Quantum Systems}, \orgname{University of Basel}, \orgaddress{\street{Klingelbergstrasse 82}, \city{Basel}, \postcode{CH-4056}, \country{Switzerland}}}




\abstract{The two-dimensional electron gas (2DEG) in reduced strontium titanate offers a versatile platform for oxide electronics, yet its dissipation mechanisms under field driven charge fluctuations remain poorly understood. Here, we combine low-temperature atomic force microscopy with scanning tunneling spectroscopy to probe the force and dissipation responses of a mechanical oscillator interacting with the STO 2DEG. The observation of Rydberg-like image potential states by tunneling experiments confirm the 2DEG formation, while dissipation spectroscopy reveals bias-dependent peaks linked to local electrostatic gating and charge redistribution within the 2DEG energy sub-bands. These features are quantitatively explained by variations in quantum capacitance as carrier density is tuned by electric fields. Under magnetic fields, dissipation peaks obey the Kohler’s rule, allowing extraction of carrier mobilities in each sub-band. Our results establish a non-invasive AFM-based methodology for quantifying energy losses in quantum oxides, providing new insights into charge dynamics relevant for spintronic applications.
}





\maketitle
\section{Introduction}
 Strontium titanate SrTiO$_3$(001) (STO), a  semiconductor with a wide electronic band gap ($\sim$ 3.2 eV),  is a paradigm example of quantum paraelectric perovskite oxide \cite{Mueller1979}, exhibiting a high dielectric constant at low temperatures without undergoing a ferroelectric transition. Absent in the pristine form of STO due to quantum fluctuations, a long-range ferroelectric order can be introduced through chemical doping, stress \cite{Uwe1976,Zhang2015}, electric fields \cite{Fleury1968} or oxygen vacancy engineering \cite{Bickel1989}. When strongly reduced \cite{Santander2011,Santander2014,Sokolovic2025,Meevasana2011,Plumb2014,Ohta2007} or interfaced with LaAlO$_3$ \cite{Annadi2013,Ohtomo2004,Taniuchi2016}, the interfacial band bending can lead to the formation of a two-dimensional electron gas (2DEG) near its surface without affecting its bulk insulating property \cite{Coey2016}.  The 2DEG yields a series of gate-tunable quantum phenomena including insulator to metal, insulator to superconductor transitions  \cite{Pavlenko2013,Schooley1964}, magnetism \cite{King2014,Santander2014}, and Rashba interaction \cite{Caviglia2010,Fete2014}.  STO has thus garnered a significant interest for the diverse physical properties emerging at its interface controllable with external stimuli \cite{Li2019}.  This intimate connection between the STO atomic configuration, its electronic properties and exotic electrostatic response makes this material a prime candidate for novel electronic applications in nanostructured materials. However, the atomic scale understanding of how energy may dissipate when fine tuning the 2DEG occupancy is lacking in the literature \cite{Cen2008,Cen2009}.
 
The plethora of oxygen-deficient STO reconstructions depends on subtle stoichiometric variations at the surface introduced by thermal treatment or cleaving procedures \cite{Gerhold2014,Dagdeviren2016}. Annealing STO at $T$ = 1050 °C in ultra-high vacuum conditions (UHV) leads to the oxygen-deficient $\sqrt{5} \times \sqrt{5}$ reconstruction, consisting of a lattice of oxygen vacancies $V_{\rm O}$  \cite{Tanaka1993}, around which each dangling bond with neighboring Ti atoms hosts two electrons. First-principles calculations showed that the $V_{\rm O}$ complexes act as an effective quantum dot \cite{Kisiel2018a}, which induces localized states in the semiconducting gap of STO \cite{Brovko2017}. At high density, the unpaired electrons of the vacancies delocalize and contribute to the formation of the surface 2DEG   \cite{Santander2014,Santander2011,Meevasana2011,Plumb2014, DiCapua2012, Guedes2021,Ohta2007,Taniuchi2016}, leading to ferroelectricity by the formation of polar nanodomains and ferromagnetism \cite{Taniuchi2016,Altmeyer2016,Santander2014}. Angle-resolved photo-electron spectroscopy (ARPES) has assigned to the 2DEG two light electron sub-bands (about -180 meV and -90 meV)  and a heavy sub-band ($\approx$  -40 meV) below $E_{\rm F}$, as a result of a surface confinement by an inherent field on the order of 10$^8$ Vm$^{-1}$ (0.1 V nm$^{-1}$) \cite{Santander2011,Sokolovic2025}. In this context, it is reasonable to assume that the tip of an AFM used as a local gate, applying voltage of few volts across a tip-sample gap of few nanometres, may electrostatically vary the carrier density within the STO's 2DEG and generate energy dissipation at the local scale.

Atomic force microscopy (AFM) operated in non-contact mode is a powerful tool to not only visualize in real space oxide surfaces \cite{Huetner2024,Sokolovic2019}, but also quantify extremely delicate forces between tip and sample \cite{Kisiel2018a,Kisiel2015} with respect to the local electronic density of states (LDOS) of the sample. Acting as a local gate, the AFM technique combined with scanning tunneling microscopy (STM) has enabled single-electron charging in individual QDs \cite{Stomp2005}, the control over the charge-state of adatoms \cite{Gross2009}, vacancies \cite{Setvin2017,Kisiel2018a} and single molecules \cite{Kocic2015,Fatayer2019,Li2023}. Through capacitive coupling between the tip's electric field and the surface,  depletion or accumulation of carriers into states located near the Fermi level $E_{\rm F}$ can be locally controlled, yielding substantial energy dissipation $\Gamma$ of the oscillator when the coupling with the substrate is due to non-conservative interaction forces \cite{Scheuerer2020,Dastolfo2022,Ollier2023,Kisiel2018a}. Combined with tunneling spectroscopy, the position in energy of the gated electronic states can be also determined with respect to $E_{\rm F}$. Using field-effect resonant tunneling (FERT) spectra, 
image potential states (IPS) lying above the vacuum level can be also probed via the tip-induced electric field \cite{Dose1987,Wahl2003,Yildiz2019,Chahib2024}. The spectra reveal a Rydberg series of surface image states bound by the crystal image potential - a hallmark of a surface 2DEG.

\begin{figure*}[t!]
	\centering
	\includegraphics[width=0.95\textwidth]{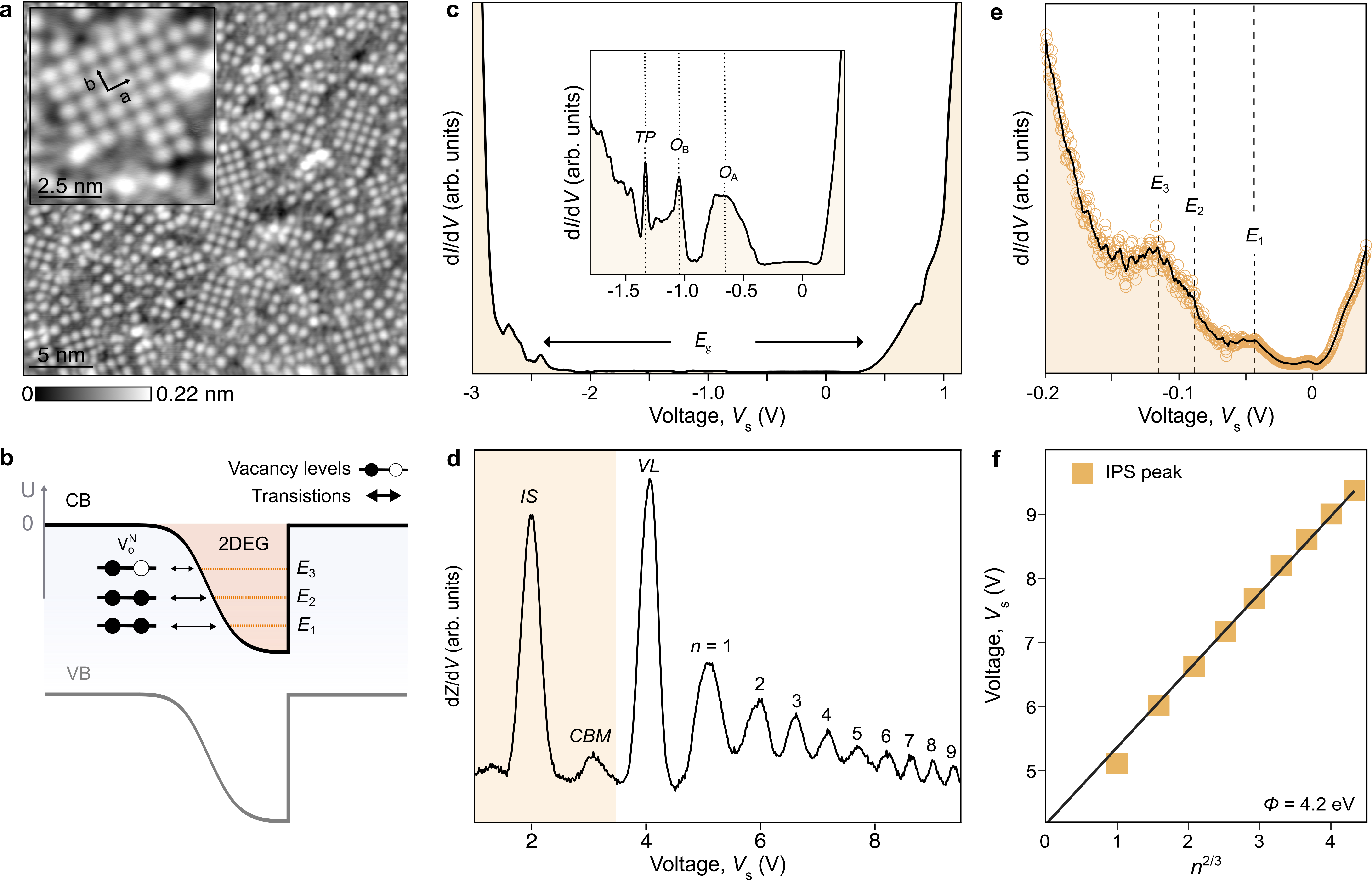}
	\caption{\textbf{Electronic properties of reduced STO surface.}
	\textbf{a}, Overview STM image of the reduced STO surface after annealing at $T$ = 1050$^\circ$C, ($I_{\rm t}$ = 1 pA, $V_{\rm s}$ = 1V) with inset showing a close-up STM image of the STO surface with atomic resolution. 
	\textbf{b}, Electron exchange between vacancies and surface 2DEG caused by the potential of an oscillating AFM tip.
	\textbf{c}, d$I$/d$V$ point-spectra showing the electronic band gap $E_{\rm g}$. The inset shows 
	 "Tanaka" peak ($TP$) characteristic of the oxygen-deficient STO surface, whereas $O_{\rm A}$ and O$_{B}$ are vacancy states.
	\textbf{d}, Single FERT spectra of the reduced STO surface showing the series of $n^{\rm th}$ IPS ($I_{\rm t}$ = 1 pA, $A_{\rm mod}$ = 50 mV, $f_{\rm mod}$ = 611 Hz). $IS$, $CBM$ and $VL$ refer to an interface state, the conduction band maxima and vacuum level, respectively. 
	\textbf{e}, d$I$/d$V$ spectrum of the reduced surface taken close to Fermi energy is characterized by the presence of three filled surface states localized at $E_{\rm 1}$ at $\sim$ -40  meV, $E_{\rm 2}$ $\sim$ -87 meV and $E_{\rm 3}$ $\sim$ -112 meV.
	\textbf{f}, Extracted IPS peak voltages as a function of $n^{\rm 2/3}$ allowing an estimate of the work function $\phi$ = 4.2 eV.}
	\label{fig:a}
\end{figure*}

In this work, we detect the force and dissipative responses of a mechanical oscillator coupled to the 2DEG of reduced STO under external electric and magnetic fields. Using tunneling spectroscopy, we characterize the local density of states induced by oxygen-vacancies at the reduced STO surface as well as the three 2DEG sub-bands near $E_{\rm F}$. The metallic nature of the STO's 2DEG and its electrostatic response to the tip field is further confirmed by probing well-defined IPS states in FERT spectra. At larger distances of the order of few nanometers, force spectroscopy enables to deplete or accumulate carriers within the 2DEG by capacitive coupling, leading to prominent force (dissipation) kinks (peaks). By applying a normal magnetic field, the force/dissipation response as a function of measured force evolve in a quadratic manner 
(in agreement with Kohler's rule of magneto-resistance \cite{he_impurity_2011, Yildiz2019}), which allows an estimate of the electron mobilities of the 2DEG. At $B=$ 0.43 T, a discontinuity appears in the force/dissipation spectra, suggesting spin polarization of the 2DEG heavy sub-band \cite{King2014}. Dissipation spectroscopy thus offers an efficient and completely non-invasive way for analyzing gate-tunable metal oxide interfaces, of considerable importance for electrically and magnetically tunable spintronic functionalities.

\section*{Results}
\subsection*{Local spectroscopic signature of the 2DEG}\label{subsec1}
Figure \ref{fig:a}a shows an overview STM image of monoatomic steps and large terraces without any sign of contamination (see Figure S1a) and it resolves the surface atomic configuration, which consists of small patches of $\sqrt{5}\times\sqrt{5}$ domains (see inset). The square lattice formed by the bright dots with parameters $a$ = $b$ = 0.86 $\pm$ 0.02 nm are found in agreement with Ref. \cite{Tanaka1993}, whereas the dark features between domains are assigned to oxygen vacancies $V_{\rm O}$, which might lead to one or several consecutive electron transitions, taking place by resonant tunneling or generally by electron transfer between vacancies and surface 2DEG \cite{Kisiel2018a, Brovko2017} as shown in Figure \ref{fig:a}b. 

Despite the large amounts of vacancies, the STO surface retains a large band gap $E_{\rm g}$ = 3.25 eV as determined by the differential conductance d$I$/d$V$ measurements of Figure \ref{fig:a}c. The valence band maximum (VBM) and conduction band minimum (CBM) are respectively located at -3 eV and 0.5 eV similar to Refs. \cite{takeyasu_control_2013,Kisiel2018a}. 
In the inset of the Figure \ref{fig:a}c, the d$I$/d$V$ spectra shows three resonances denoted as $O_{\rm A}$, $O_{\rm B}$ and $TP$, which are attributed to occupied electronic states  from oxygen vacancies as described in Ref. \cite{Tanaka1993, Kisiel2018a}. At about -0.7 eV below $E_{\rm F}$, $O_{\rm A}$ is the shallower in-gap state due to the oxygen vacancy $V_{\rm O}$ state which occupancy can be controlled by external tuning of chemical potential \cite{Kisiel2018a}. At about -1.3 eV and -1.1 eV, the "Tanaka" peak $TP$ and  $O_{\rm B}$ are also characteristic signatures of the strongly reduced STO surface. 

Let us now discuss the 2DEG of the reduced STO surface. To compare with previous ARPES results \cite{Santander2014,King2014}, we characterize the DOS near $E_{\rm F}$ using d$I$/d$V$ spectra (Figure \ref{fig:a}e). Composed of three filled surface states, the 2DEG signature is attributed in our data to the shoulders $E_{\rm 1}$ at $\sim$ -40  meV, $E_{\rm 2}$ $\sim$ -87 meV and $E_{\rm 3}$ $\sim$ -112 meV, respectively.  Their respective positions in energy agree with  previous ARPES data \cite{Santander2014,King2014}, allowing to conclude that $E_{\rm 1}$ is the heavy carriers sub-band with effective mass $m^*$ = 10 $m_{\rm e}$ ($m_{\rm e}$ is the free electron mass) whereas $E_{\rm 2}$ and $E_{\rm 3}$ may correspond to light carriers with effective masses $m^*$ = 0.7 $m_{\rm e}$. We also stress that the presence of those states has been reproduced on many reconstructed patches of the surface (Figure S2).  

To unambiguously confirm the 2DEG's presence, we also performed FERT spectroscopy at the vicinity of the surface (Figure \ref{fig:a}d). The IPS originating from the Rydberg series of surface states confined between the surface and crystal image potential are the hallmark of surface states (i.e. 2DEG) at metallic surfaces \cite{Dose1987,Wahl2003}. Experimentally, we acquired d$Z$/d$V$ spectroscopic measurements near the STO surface by a continuous sweep of the tip-sample voltage $V_{\rm s}$ while keeping constant the tunneling current with the STM feedback loop (see Methods). When $V_{\rm s}$ exceeds the local work function ($\phi$), resonant tunneling between tip and sample occurs as soon as the Fermi level of the tip aligns with the IPS states. This leads to the series of peaks in the FERT spectra as exemplary shown in Figure \ref{fig:a}d, which position can be described from a quasi-classical approximation by the equation: 
\begin{equation}
\textrm{e} V_{\rm n} = \phi + \left( \frac{3n \uppi \uphbar \textrm{e} E}{\sqrt{2m_{\rm e}}}\right)^{^{^{2/3}}}
\label{eq1}
\end{equation}
where $V_{\rm n}$ is the sample voltage for the $n^{\rm th}$ IPS, $\phi$ is the local work function of the sample, $m_{\rm e}$ is the free electron mass and $E$ is the electric field. 

In Figure~\ref{fig:a}d, the IPS resonances are marked with their index numbers $n$ = 1..9 (see details in Supplementary Information S3). Note also that the IPS oscillations in FERT spectra are homogeneous across the surface, meaning that we observed no, or minor alterations near defects and domain boundaries (Fig. S3b,c). IPS are also completely absent on pristine (non-reduced) STO surface (Fig. S3a in Supplementary Information S3). The resonance denoted $IS$ corresponds to an interfacial state, whereas $CBM$ and $VL$ are the conduction band maximum and vacuum level, respectively. A quantitative estimation of the local work function of the sample is obtained from Eq.~\ref{eq1} by plotting  the voltage position $V_{\rm n}$ of the IPS states as a function of $n^{2/3}$ (orange squares in Fig.~\ref{fig:a}f). By fitting the linear regression, we extract the $\phi$ value corresponding to the $y$-intercept equal to about  4.2 eV, a value in good agreement with previous works \cite{Chung1979}. Altogether, the observation of oxygen-vacancy states and IPS in tunneling spectra confirms the emergence of the 2DEG in our sample.

\subsection*{Dissipation spectroscopy under electric field}\label{subsec2}
\begin{figure*}[h]
			\centering
			\includegraphics[width=0.95\textwidth]{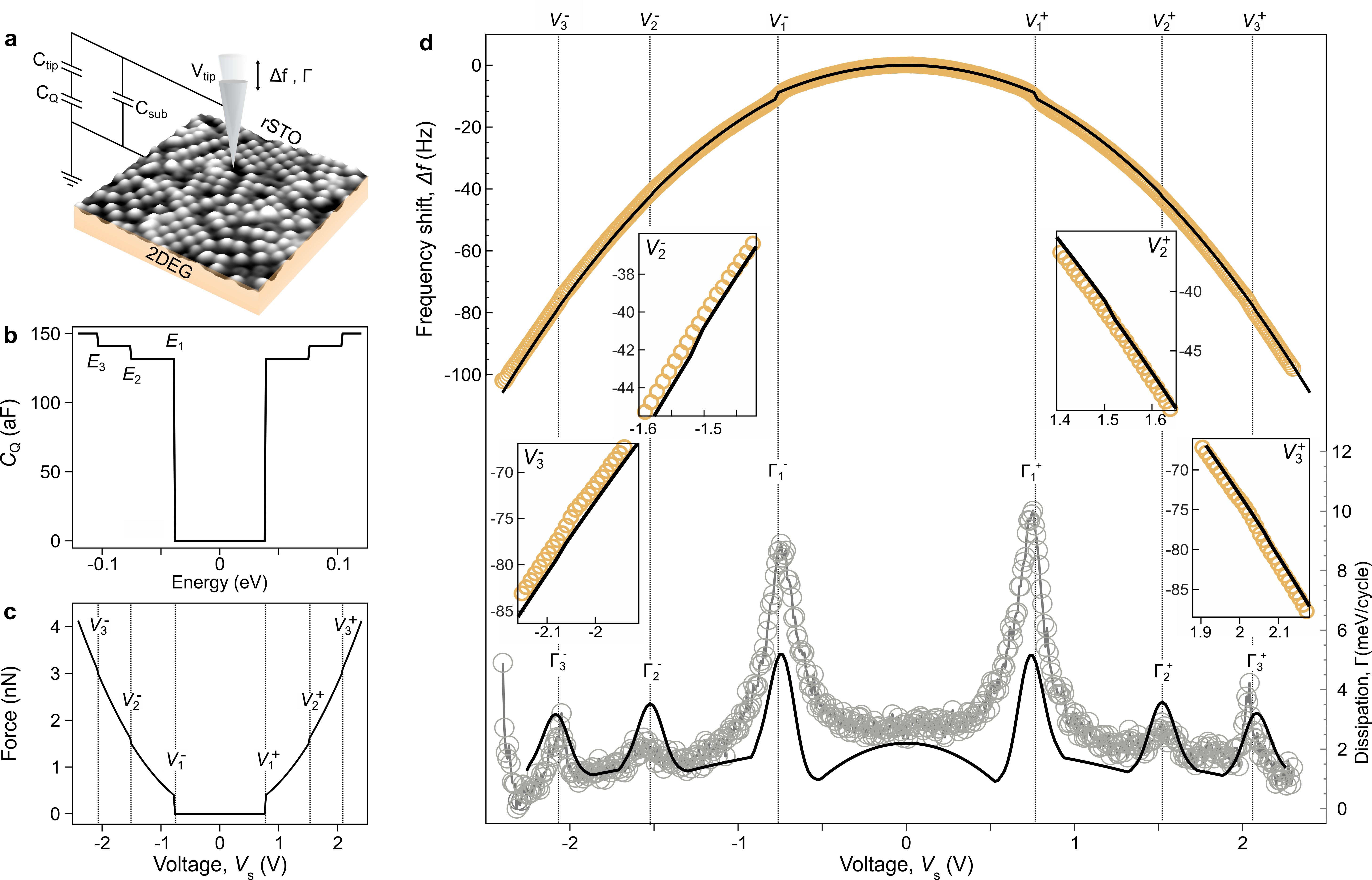}
			\caption{\textbf{Force and dissipation spectroscopy of the  2DEG.}
			\textbf{a}, Principle of the experiment : the AFM tip is capacitively coupled to the 2DEG of reduced STO surface. $C_{\rm tip}$ and $C_{\rm sub}$ refer to the capacitance of the tip and substrate, respectively. $C_{\rm Q}$ is the 2DEG quantum capacitance. 
			\textbf{b}, Quantum capacitance $C_{\rm Q}$ as function of energy. $C_{\rm Q}(E)$ is modeled by a Heaviside function, whose step positions and heights derive from the energy of the 2DEG sub-bands $E_{\rm 1, 2, 3}$ and their effective masses $m^*_{\rm 1, 2, 3}$.
			 \textbf{c}, Tip-sample force versus voltage $F(V)$ curve simulated from the $C_{\rm Q}$ function, whose steps induces the $V^{\pm}_{\rm 1,2,3}$ occurrences (dashed lines). 
			 \textbf{d}, Experimental $\Delta f(V)$  (orange) and dissipation $\Gamma(V)$ (gray) spectra. Three kinks in the $\Delta f(V)$ parabola are observed at voltages $V^{\pm}_{\rm 1,2,3}$ and accompanied by pronounced dissipation peaks $\Gamma^{\pm}_{\rm 1,2,3}$.  Black spectrum corresponds to simulated  $\Delta f(V)$ and dissipation  $\Gamma(V)$ from the $F(V)$ curve of {\bf c}. The experimental spectra are corrected by the contact potential difference $V_{\rm CPD}$ = -1.2 V by lateral shift.}
			\label{fig:b}
		\end{figure*}
External perturbations caused by the local electric field of an oscillating AFM tip can push a quantum system towards a transition or a level crossing with subsequent fluctuation and relaxation of the system, eventually resulting in energy losses \cite{Kisiel2015,Kisiel2018a,Yildiz2019,Dastolfo2022}.  With the tip oscillating at distance Z of about 5 nanometer above the STO surface, we measure the frequency shift $\Delta f$ and mechanical dissipation $\Gamma$ of the qPlus AFM (see Methods) coupled to the STO's 2DEG while varying the tip-sample voltage $V_{\rm s}$.  Typical experimental $\Delta f(V)$ (orange) and $\Gamma (V)$ (gray) spectra, for which we have corrected the contact potential difference (CPD), are shown in Figure~\ref{fig:b}d. Three kinks are observed along the $\Delta f(V)$ parabola at $V^{\pm}_{\rm 1,2,3}$ voltages (highlighted in the insets of Fig.~\ref{fig:b}d), which are accompanied by pronounced dissipation peaks denoted $\Gamma^{\pm}_{\rm 1,2,3}$. Each dissipation peak is characterized by a dissipated power on the order of few meV per cycle, which is a value typical for single-electron charging of quantum dots \cite{Stomp2005, Kisiel2018a,Dastolfo2022}.  

To better understand the origin of the mechanical dissipation, we model the system schematized in Figure~\ref{fig:b}a with three capacitances ($C_{\rm tip}$, $C_{\rm sub}$ and $C_{\rm Q}$), which represents the geometrical tip-sample capacitance, the substrate (background) capacitance and the quantum capacitance exerted by the 2DEG, respectively. While $C_{\rm tip}$ and $C_{\rm sub}$ are well-established to describe the parabolic lineshape of $\Delta f(V)$ spectroscopic measurements,  $C_{\rm Q}$ reflects how the 2DEG localized at the vicinity of the STO surface can accommodate additional charges at a given quantum level \cite{Luryi1988}. This capacitance expresses as $C_{\rm Q}$ = $\rho e^2$ = $\frac{m^*e^2}{\pi \hbar^2}$ with  $\rho$, $e$ and $m^*$ are the electronic DOS, the elementary charge and the effective mass of the quantum state, respectively.  For a 2DEG, the density of states as a function of energy $\rho(E)$ can be approximated with a Heaviside step function, whose step positions correspond to the energies of the 2DEG sub-bands $E_1$, $E_2$ and $E_3$ and step heights directly derive from the in-plane effective masses of the corresponding sub-bands. Since $E_1$ is the heaviest band as compared to $E_2$ and $E_3$,  the $C_{\rm Q}(E)$ $\propto$ $\rho (E)$ function, shown in Figure \ref{fig:b}b has a large step associated to $E_1$ and two smaller and equal steps for $E_2$ and $E_3$, respectively.

To account for the interaction forces acting on the cantilever, the electrostatic force between tip and sample plotted in Figure \ref{fig:b}c is obtained by considering $C_{\rm Q}(E)$ in the equation introduced by Stomp et al. \cite{Stomp2005}. By derivation, we simulated the $\Delta f(V)$ and dissipation spectra such as $\Gamma(V) \propto V \frac{\partial (\Delta f)}{\partial V}$, which are superimposed on the experimental spectra (black curves of Fig. \ref{fig:b}d). Details of the simulation can be found in Supplementary Information S4. Note that, for these simulations, the effective tip radius $R$ = 26 nm is the only free parameter. The coupling between tip and sample background capacitance $C_{\rm sub}$ leads to the parabolic shape of $\Delta f(V)$ spectra, whereas the steps of the curve are induced by the response of the 2DEG to the electric field of the AFM tip. This response is capacitive by nature and thus related to the lever arm $\alpha = \frac{C_{\rm {tip}}}{C_{\rm {tip}}+<C_{\rm Q}>}$ = 0.056 (i.e. the gating efficiency between tip and sample). As a result,  the voltage thresholds $V_{\rm 1,2,3}^{\pm}$ corresponding to the $\Delta f^{\pm}_{\rm 1,2,3}$ drops and $\Gamma^{\pm}_{\rm 1,2,3}$ dissipation peaks (dashed lines of Fig.~\ref{fig:b}d) scales as  $V_{\rm 1}$ = $E_1 / e\alpha \approx$ 0.7 V, $V_{\rm 2}$ = $E_2 / e\alpha \approx$ 1.5 V and  $V_{\rm 3}$ = $E_3 / e\alpha \approx$ 2.0 V, respectively. The largest $\Delta f$ drop ($\Gamma$ peak) is observed for  $V_{\rm 1}$ due to the heavier mass of the $E_{\rm 1}$ sub-band as compared to $E_{\rm 2}$ and $E_{\rm 3}$ . 

Since the model of Figure \ref{fig:b}a successfully reproduces the $\Gamma^{\pm}_{\rm 1,2,3}$ peaks as well as their $V^{\pm}_{\rm 1,2,3}$ positions, we also considered for a better agreement a broadening effect by taking into account a tunneling rate of the gating process equal to $\gamma = 4\pi^2kA^2\frac{\delta f}{E_{\rm diss}} \approx$ 1.2 MHz.  This value is one order of magnitude larger as compared to the tunneling rate measured in the confined 2DEG on Ag(111) surface \cite{Dastolfo2022} (Fig. S4 in Supplementary Material S5). Note that presence of $C_{\rm sub}$ imposes parabolic background in $\Delta f(V)$ signal, whereas $\Gamma (V)$ shows no such parabolic Joule-dissipation component \cite{Kisiel2011} and we consider that in the simulations. Its absence is due to small (100pm) tuning fork oscillations, perpendicular to the sample surface, which explains that Joule looses are not detected - point we revisit later when comparing with spatially extended pendulum AFM oscillations.

\begin{figure*}[t!]
			\centering
			\includegraphics[width=0.95\textwidth]{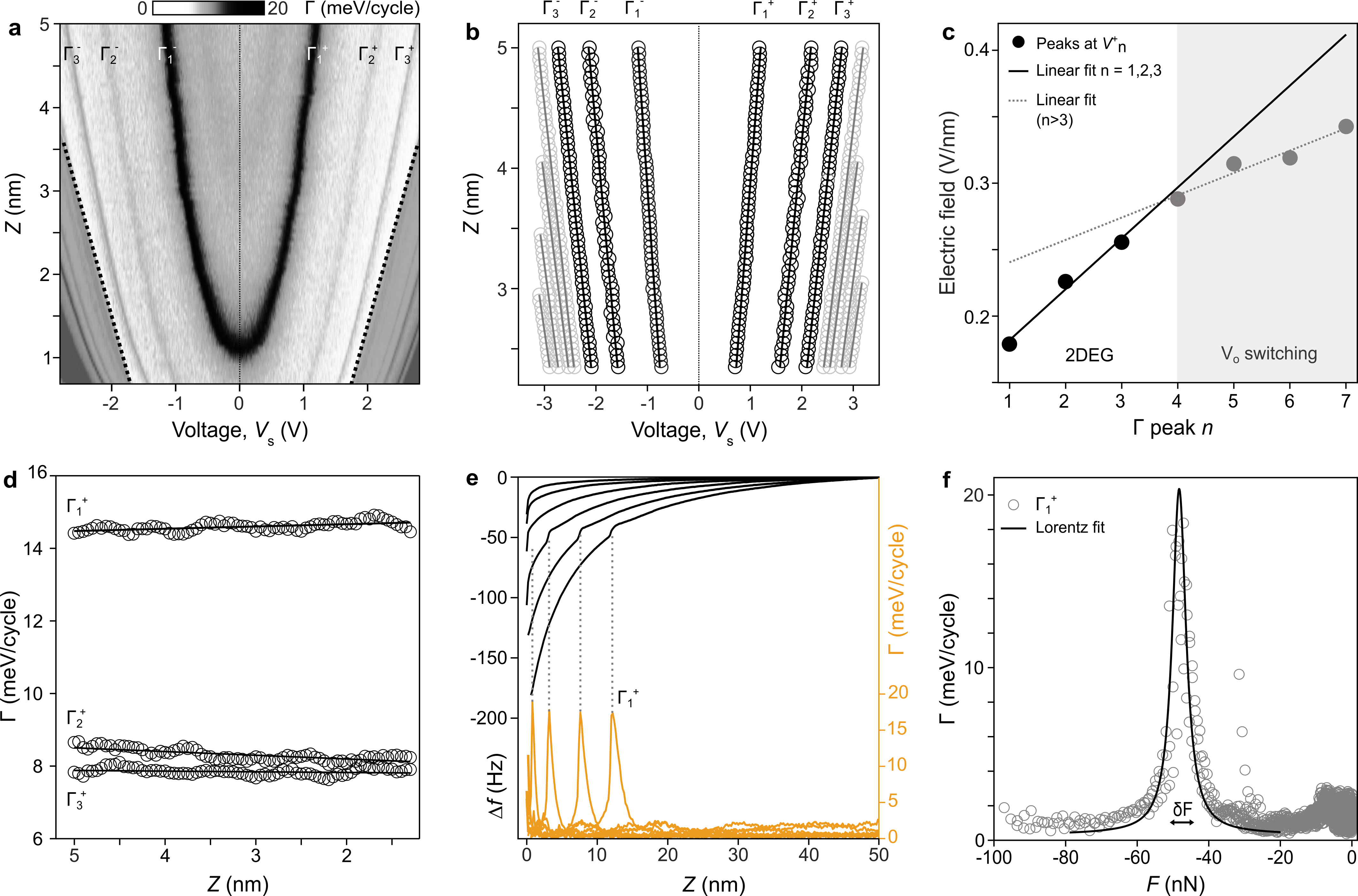}
			\caption{\textbf{Evolution of dissipation peaks as a function of tip-sample distance $Z$.} 
			\textbf{a}, $\Gamma (V,Z)$ dissipation map between the reduced STO surface and the AFM tip. The $V_{\rm CPD}$ value has been corrected for all dataset. Dark lines denoted $\Gamma^{\pm}_{1,2,3}$ are the dissipation peaks shown in Fig. \ref{fig:b}. The gray area for large voltages and/or close distances is attributed to tip-induced switching of the $V_{\rm o}$ states.
			\textbf{b}, Linear regime of the $\Gamma(V,Z)$ map shown in \textbf{a} used to extract the slope (tip-sample electric field strength) for each dissipation peak.
			\textbf{c}, extracted electric fields for dissipation peak marked by $n$ separate two different charging mechanisms:  2DEG-induced charging peaks (white region) and -$V_{\rm o}$ vacancy induced charging (grey), distinguished by their slopes.
			\textbf{d}, dissipation peak intensities versus distance show that $\Gamma(Z)$ is constant for each $\rm n=$1..3 peak, with the largest value of about 14meV per cycle for $\Gamma^{\pm}_{1}$ dissipation peak.
			\textbf{e}, Series of $\Delta f(Z)$ spectra (black) and corresponding $\Gamma(Z)$ (orange) acquired for different $V_{\rm s}$. The step in $\Delta f(Z)$ (dashed lines) corresponds to the $\Gamma^{\pm}_1$ dissipation peak. 
			\textbf{f}, Extracted dissipation versus force $\Gamma (F)$ spectra for different bias voltages $V_{\rm s}$. All dissipation peaks emerge at constant attractive force interaction $F$ = 50 nN, independently of the applied voltage $V_{\rm s}$ value. Lorentzian fit is overlaid as a guide to the eye.}
			\label{fig:c}
\end{figure*}

The dissipation map $\Gamma (V,Z)$ of Fig. \ref{fig:c}a shows the evolution of the $\Gamma^{\pm}_{\rm n}$ peaks ($n$ is an index) and their voltage thresholds $V^{\pm}_{\rm n}$ as a function of the tip-sample distance $Z$. The spectrum at the largest tip-sample distance $Z$ = 5 nm corresponds to that of Figure \ref{fig:b}, where the most prominent intensity lines is $\Gamma^{\pm}_1$ followed by the fainter $\Gamma^{\pm}_{2,3}$ peaks. Their contrasts are due to the difference of magnitude between them as shown in the spectra of Fig. \ref{fig:b}d. In the gray areas of Fig.~\ref{fig:c}a, additional dissipation peaks are detected for close tip sample distances and/or high voltage values, which will be discussed below.  The $\Gamma^{\pm}_1$ branch clearly exhibits a parabolic behavior at close tip-sample distance ($Z \leq$ 2.5 nm in Fig.~\ref{fig:c}a), which results from the capacitive force $F \propto \frac{V^2}{Z}$ arising between a conical tip and the flat sample. For simplicity, we will thus limit the following discussion to tip-sample distances greater than $Z \geqslant $ 2.5 nm, where the capacitive response between the tip and the sample is purely linear (Fig.~\ref{fig:c}b). 

This linear regime of the experimental $\Gamma^{\pm}_{\rm n}$ occurrences versus $Z$ is plotted in Fig.~\ref{fig:c}b, where black dots are the $\Gamma^{\pm}_{1,2,3}$ occurrences and gray dots the additional $\Gamma^{\pm}_{n > 3}$. The extracted slopes $\eta$ (black lines of Fig.~\ref{fig:c}b) are the lever arm gradient $\eta$ = e$V_{\rm s}\frac{d\alpha}{dZ}$ for each $\Gamma^{\pm}_{\rm n}$ dataset, which corresponds to the strength of the electric field $\tilde{E}$ between tip and sample. In Fig.~\ref{fig:c}c, the $\eta$ values are plotted as a function of the index $n$, allowing us to extract two linear regressions (black and gray dotted lines of Fig.~\ref{fig:c}c), which scales as $\Delta \tilde{E}$ = $\frac{\Delta \rm N \cdot e}{2\pi\varepsilon_0 \varepsilon_r \rm R^2}$, where $\Delta N \cdot e$ is the charge difference between subsequent charging peaks and $\varepsilon_0$, $\varepsilon_r$ are vacuum and sample permittivities, respectively. It indicates that the change of the electric field $\Delta \tilde{E}$ for $n$ = 1,2,3  is similar upon charging/discharging, but differs from the $\Gamma^{\pm}_{n > 3}$ peaks. Since  $\Delta \tilde{E} \propto \frac{\Delta N \cdot e}{\varepsilon_r}$,  the  polarization of electric charges induced by the tip field leads to two distinct dissipation responses. We attribute these two categories to $i$- the 2DEG response for $n \leqslant$ 3 and $ii$- charge-state transitions of individual or possibly collective groups of oxygen vacancies $V_{\rm o}$ for $n >$ 3, as we reported previously \cite{Kisiel2018a}. Note that, although tunneling in IPS can also open dissipation channels \cite{Yildiz2019}, we do not consider it as candidate for explaining the $\Gamma^{\pm}_{n > 3}$ peaks since the employed voltages are much smaller than the vacuum level of the sample (Fig.~\ref{fig:a}d).  

The field-induced charge fluctuations of the 2DEG result in energy losses, which magnitudes can be estimated by comparing the $\Gamma^{+}_{\rm 1,2,3}$ peak values as a function of $Z$ (Fig.~\ref{fig:c}d). The strongest dissipation of 14-15 meV/cycle is obtained for the $\Gamma^{\pm}_{\rm 1}$ peak, a value twice larger than the $\Gamma^{\pm}_{\rm 2,3}$ peaks (8-9 meV/cycle), and remains constant versus $Z$. This not only suggests that the dissipation peak consistently occurs at constant tip-sample interaction force independent of $V_{\rm s}$, but also indicates that the underlying mechanism for dissipation is governed by a constant force (electric field) rather than the voltage value.  

A quantitative estimate of the interaction force can be extracted from the series of $\Delta f(Z)$ (black) and corresponding $\Gamma(Z)$ (orange) spectra acquired for voltages $V_{\rm s}$ (Figure \ref{fig:c}e).  Independent on the applied voltage, all spectra shows a pronounced kink at $\Delta f$ = 42 Hz accompanied by the $\Gamma_1$ dissipation peak. The corresponding interaction force was extracted via the Sader-Jarvis inversion algorithm \cite{Sader2004}. 

By displaying the dissipation as a function of force for all $V_{\rm s}$-dependent $F(Z)$ (Fig.~\ref{fig:c}f), tuning-fork systematically show reproducible dissipation maximum positioned at single value of attractive force of approximately $F=$ -50 nN. This indicates that charge transfer into the 2DEG is set by a constant tip-sample force (or electric field), not by the applied voltage. Thus, the effect is force-controlled rather than voltage-controlled.

\subsection*{Dissipation spectroscopy under magnetic field}\label{subsec3}
\begin{figure*}[t!]
	\centering
	\includegraphics[width=0.9\textwidth]{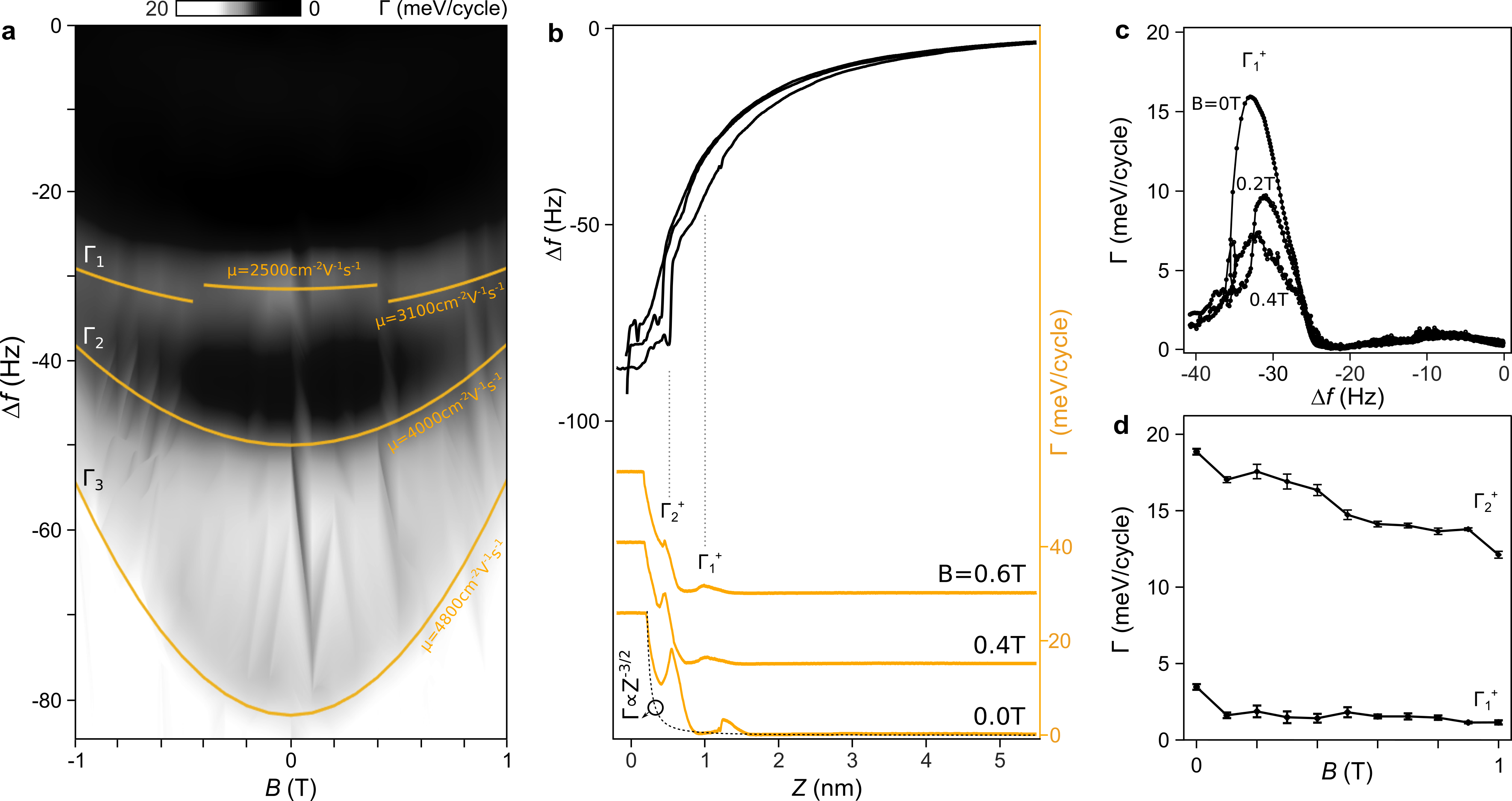}
	\caption{\textbf{Pendulum-AFM energy dissipation vs. magnetic field. }
	\textbf{a}, dissipation intensity map as a function of $\Delta f$ and external magnetic field $B$. For specific dissipation peaks $\Gamma_1$, $\Gamma_2$ and $\Gamma_3$, $\Delta f(B)$ follows parabolic dependence (orange guides) consistent with standard Kohler's rule, which enables extraction electron mobilitiy $\mu$ for each sub-band. A discontinuity is visible for heavy $\Gamma_1$ sub-band, suggesting possible magnetic polarization (see text). 
	\textbf{b}, Series of $\Delta f(Z)$ spectra (black) and corresponding $\Gamma(Z)$ (orange) acquired for different $B$-field. The steps in $\Delta f(Z)$ (dashed lines) corresponds to the $\Gamma^{\pm}_1$, $\Gamma^{\pm}_2$ dissipation peaks. The contact potential $V_{\rm CPD}$ was compensated to minimize the influence of electrostatic forces. 
	\textbf{c}, Extracted $\Gamma^{\pm}_1(\Delta f)$ spectra for various magnetic fields $B$. The peak shifts with $B$-field: at $B$=0.2T it moves to lower $\Delta f$ values, whereas at $B$ = 0.4T it shifts to higher $\Delta f$ and broadens. This behavior is consistent with the dissipation map shown in \textbf{a}.
	\textbf{d}, dissipation peak intensities versus $B$-field show that $\Gamma_1(B)$ is roughly constant at ~2 meV/cycle, whereas $\Gamma_2(B)$ decreases approximately linearly from 18 to 12 meV/cycle between $B$ = 0 and 1T.}
	\label{fig:d}
\end{figure*}

The further characterization is continued with an external magnetic field $B$, where we measure $\Gamma(B)$ dependence. Because magnetic forces are much weaker than competing electrostatic forces, we minimize the latter by setting the bias equal to contact potential difference $V$ = $V_{\rm CPD}$, and 
we performed the experiment on an alternative instrument, namely the pendulum AFM (pAFM). Owing to low spring constant and high quality factor, pAFM provides significantly enhanced force and dissipation sensitivity as compared to the tuning fork. However the in plane tip oscillations - parallel to the sample surface - leads to increased spatial delocalization of the measurements and non-trivial derivation of the force fields (see Methods and Supplementary Material S1). The force distance curves were acquired as in Figure \ref{fig:c}e, except that - instead of varying the voltage - we change out-of-plane $B$-field from -1T to 1T. Few of pAFM $\Delta f(Z)$ (black) and corresponding $\Gamma(Z)$ (orange) spectra acquired for selected, positive $B$-fields are shown in Figure \ref{fig:d} b. Independent on the $B$-field, all spectra show kinks marked with dashed lines accompanied by the $\Gamma_1$, $\Gamma_2$ dissipation peaks. Unlike the tuning-fork data (see Figure \ref{fig:c}e), the pAFM dissipation signal - beside these peaks - exhibits a power-law distance dependence versus distance. At $B$ = 0T, the dashed curve indicates $\Gamma \propto Z^{-\frac{3}{2}}$ background - a scaling that agrees with the theory of non-contact dissipation in a 2D electron system on an insulator \cite{volokitin_giant_2007}. A parabolic, electrostatic background is also present/absent in pAFM/tuning fork dissipation versus voltage data (Fig. S5 in Supplementary Information S6). The difference is most likely due to lateral oscillations of pAFM tip, whereas the tuning-fork normal oscillations are more local \cite{volokitin_giant_2007}. Figure \ref{fig:d}c plots $\Gamma^{+}_1$ versus $\Delta f$ for few $B$-fields. Owing to pAFM in-plane oscillation and resulting complex $F$-field, we could not convert $\Delta f$ to $F$ as in Figure \ref{fig:c}f, however the small normal oscillation amplitudes encode the same physics: $\Delta f \propto \frac{\partial F}{\partial z}$. $\Gamma^{+}_1(\Delta f)$ position depends on $B$ - unlike in the electric field case (Figure \ref{fig:c}f), where it was constant at a given $F$, independently on applied bias voltage. From $B$ = 0T to 0.2T the peak shifts to lower $\Delta f$ and at $B$ = 0.4T it moves back to higher $\Delta f$ and broadens. The overall peaks intensity also decreases (Fig. \ref{fig:d}d). The two peaks $\Gamma^{+}_1$ and $\Gamma^{+}_2$ are unequal in intensity, yet as discussed before, their heights include the electrostatic background. The full dependence of dissipation versus B-field and $\Delta f$ is shown in Figure \ref{fig:d}a as a intensity map (see Fig. S6 for the raw spectra). The three dissipation sub-bands $\Gamma_n$ ($n=1,2,3$) are shown and for each of them $\Delta f_n (B)$ exhibits a parabolic dependence on the magnetic field, as indicated by the solid orange lines. Notably, a tiny discontinuity is observed around B = $\pm$0.43T for heavy $\Gamma_1$ dissipation sub-band. Fitting $\Delta f(B)= a \cdot B^2+b$ to the three dissipation peaks yields $a$ and $b$ parameters for all sub-bands as shown in Table \ref{tab1}.

\begin{table}[t]
\centering
\caption{Fit parameters $a$(Hz/T$^{2}$) and $b$(Hz) for dissipation peaks as a function of magnetic field. Extracted electron mobilities $\mu$(cm$^{2}$/Vs for the 2DEG subbands.}
\label{tab1}
\setlength{\tabcolsep}{8pt}
\begin{tabular}{lccc}
\toprule
\textbf{} & $a~(Hz/T^{2})$  & $b~(Hz)$ & $\mu~(cm^2V^{-1}s^{-1})$\\
\midrule
$\Gamma_1^{\mid B \mid<0.43T}$ &  3  & -31.5 & 2500 \\  \\
$\Gamma_1^{\mid B \mid>0.43T}$  & 5 & -34 & 3100 \\
$\Gamma_2$ & 12 & -50 & 4000 \\
$\Gamma_3$ & 28 & -82 & 4800 \\
\bottomrule
\end{tabular}
\end{table}

\section*{Discussion}
Scanning tunneling spectroscopy (STS) measurements on oxygen-deficient STO reveal the presence of a metallic 2DEG evidenced by in-gap surface states ($E_1$, $E_2$, $E_3$) and field emission resonances - so called IPS - which are characteristic signatures of metallic behavior on the otherwise insulating oxide. On such surface, the electric field ($0.1\rm{Vnm^{-1}}$) introduced by non-contact AFM tip, is capable to locally modify the confinement of the 2DEG, thus providing the access to dynamic charge processes and enabling the detection of discrete charging events of the 2DEG. The presence of oxygen vacancies is essential. Their role in producing large dissipation peaks during AFM measurements has already been demonstrated in previous studies \cite{Kisiel2018a, Brovko2017}, where several charging scenarios have been proposed, including electron exchange between a single vacancy and the conduction band, between vacancies themselves, and between vacancies and the surface 2DEG (see Figure 5 in \cite{Kisiel2018a}). Our measurement, supported the latter scenario. The strong agreement between the experimentally observed charging peaks and the simulated dissipation spectra (Fig.\ref{fig:b}), which incorporate the STS energy levels of the surface 2DEG, suggests that the dominant mechanism involves electron exchange between oxygen vacancies and the 2DEG. In this scenario, the AFM dissipation response arises primarily not from the change of the vacancy's charge state, but from variations in the quantum capacitance of the de-localized 2DEG. The charge originates from vacancy states (quantum dot-like electron emitters) and the observed dissipation response is governed by the electronic compressibility or quantum capacitance $C_Q$ of the 2DEG with the vacancies acting merely as electron sources or sinks. Using the average $<C_Q>$ = 140aF (Fig. \ref{fig:b}b) and tip projected "active" surface area $S=\pi\frac{d}{2}^2$, where d = 5nm is the tip-sample distance \cite{Yildiz2019}, we can estimate a sheet capacitance $C_{2D}$ = 700$\mu \rm{F cm^{-2}}$, which for 2DEG energy width $\Delta E$ = 0.1eV (Fig. \ref{fig:a}e) yields a carrier density $N \approx 4 \cdot 10^{14} \rm{cm^{-2}}$. That places the STO 2DEG firmly in the Fermi liquid regime \cite{kim_quantum_2022}, yet many-body renormalization remains strong \cite{kingQuasiparticleDynamicsSpin2014a}. In particular, King et. al \cite{King2014} reported that pronounced orbital ordering, spin-orbit coupling and many body interactions are crucial for STO 2DEG electronic structure. The electrostatic potential that defines the 2DEG lifts the $t_{2g}$ orbital degeneracy present in the bulk STO through orbital mixing, resulting in the coexistence of light, circular $d_{xy}$ sub-bands and heavy, elliptical $d_{xz/yz}$ band. While a conventional Rashba splitting at the bottom of the $d_{xy}$ band is relatively small, the orbitally enhanced Rashba interaction can increase significantly - by an order of magnitude - near avoided crossings with the heavy $t_{xz/yz}$ band. 
Moreover, heavy $d_{xz/yz}$ sub-band couples very strongly to phonons \cite{Sokolovic2025} over a wide frequency range, resulting in substantial bandwidth renormalization, as compared to the more conventional electron-phonon (el-ph) coupling leading to kinks in the light $d_{xy}$ sub-band dispersion. Thus, the stronger el-ph interaction increases the effective mass and scattering rate of the heavy sub-band, possibly resulting in enhanced energy dissipation of $\Gamma_1$ on the order of 10-20meV per cycle, compared with few meV per cycle dissipation of the light $\Gamma_{2,3}$ (see Fig. \ref{fig:b}, \ref{fig:c}).
 
While the dissipation peaks consistently occur at the same tip-sample interaction force or electric field independently on the tip-sample voltage (Fig. \ref{fig:c}f), the dissipation behavior under the external $B$-field differs significantly. Specifically, $\Delta f$ associated with each of the $\Gamma_1$, $\Gamma_2$, and $\Gamma_3$ charging peak exhibits a parabolic dependence on $B$ (Figure \ref{fig:d}a). In a single relaxation time approximation and a weak-magnetic field limit, the magnetoresistivity follows Kohler's rule, $R(B)/R(B=0)\approx1+(\mu B)^2$ - where $\mu$ is the carrier mobility - consistent with previous results for 2DEG on liquid helium \cite{sanden_quantum_1987}. Under these assumptions, the observed $\Delta f_n (B)$ for dissipation peak marked by $n$, is expected to follow parabolic dependence (see Supplementary Material 8 for details):

\begin{equation}\label{eqn:1}
    \Delta f_n (B) \approx \Delta f_n (B=0) \cdot (1 - \frac{3}{2} \mu_n ^2 B^2)
\end{equation}

where $\Delta f_n (B=0)$ is the frequency shift at zero magnetic field. Fit of the experimental data with the formula eq.\ref{eqn:1} allows to determine the carrier mobilities corresponding to the individual $\Gamma_1$, $\Gamma_2$, and $\Gamma_3$ sub-bands, related to $E_1$, $E_2$, and $E_3$ energy in gap states. The heavy-electron $\Gamma_1$ band exhibits lower mobility values as compared to the lighter $\Gamma_2$, $\Gamma_3$. The extracted values are summarized in Table 1 and range from 2500-4800 $cm^2V^{-1}s^{-1}$, consistent with previously reported high-mobility values of an electron gas confined at the LaAlO$_3$/SrTiO$_3$ interface at low temperatures \cite{Ohtomo2004}. 

We will make one remark before closing. A tiny discontinuity is observed around $B=\pm$0.43T for the heavy $\Gamma_1$ sub-band (Figure \ref{fig:d}a), accompanied to slight increase of mobility ($\mu=\frac{e\tau}{m^*}$) for $\mid B \mid>$0.43T. Within the single relaxation time approximation, this observation implies a reduced scattering rate $1/\tau$. It is possibly due to the heavy character of $d_{xz/yz}$ band, which exhibits larger spatial delocalization as compared to its light $d_{xy}$ counterparts and consequently, its wave functions enclose more V$_o$ vacancies \cite{kingQuasiparticleDynamicsSpin2014a}. Thus, the coupling between the carriers and localized V$_o$ moments might be significant, particularly due to relatively large orbital angular momentum (0.7 $\hbar$) near the heavy-band crossings, which enhances the interaction with V$_o$ magnetic moments. Thus, at low B-fields, the random magnetic moments of individual surface vacancies V$_o$ (or oxygen-vacancy clusters) remain unpolarized and might act as fluctuating transverse magnetic fields contributing to the spin scattering events. An external B-field polarizes V$_o$ moments and once aligned they present almost no transverse component which leads to reduction of the spin-dependent scattering of 2DEG and slight increase of $\mu$ above $B=$ 0.43T.

AFM mechanical dissipation combined with STS measurements enables the characterization of electron dynamics and determination of band-selective mobility within the surface 2DEG of oxygen-deficient STO. The observed mobility change in B-field arises from suppression of scattering in surface 2DEG, possibly due to vacancy moment alignment, highlighting the 2DEG and defect interplay. The dissipation spectroscopy offers an efficient and completely non-invasive tool for analysis of correlated oxide interfaces, of considerable importance for electrically and magnetically tunable spintronic functionalities.

\section*{Methods}
\subsection*{Sample preparation}
The experiments were performed in two independent ultra high vacuum (UHV) chambers, both operating at low temperature of T = 4.5 K. A commercially available SrTiO$_3$ crystal from MTI Corporation was sputtered with Ar$^+$ ions and annealed by direct current heating to $T$ = 950 $^\circ$C as determined by an infrared pyrometer. This process leads to removal of surface contaminations and hydrocarbons from the surface. The oxygen deficient ($\sqrt{5}\times\sqrt{5}$)$R26.6^\circ$ reconstruction of SrTiO$_3$(001) is obtained by annealing at $T$ $\approx$ 1050 $^\circ$C by passing current through the sample under UHV conditions for 20 minutes.\\

\subsection*{Low temperature STM/AFM experiments}
The STM/AFM experiments were performed using a low-temperature microscope operated at $T$ = 4.8 K in ultrahigh vacuum (p $\approx$ 1 $\times$ 10$^{-10}$ mbar). The force sensor is a quartz tuning fork based on a qPlus design~\cite{Giessibl2019} operated in the frequency-modulation mode (resonance frequency $f_0$ $\approx$ 25 kHz, spring constant $k$ $\approx$ 1800 N.m$^{-1}$, quality factor $Q$ $\approx$ 14000 and oscillation amplitude $A$ $\leq$ 1~\AA). The bias voltage was applied to the tip. STM images are taken in constant-current mode. The tip mounted to the qPlus sensor consists of a 25 $\mu$m-thick tungsten wire, shortened and sharpened with a focused ion beam.  Scanning tunneling spectroscopy (STS) data acquired at low temperature with the lock-in technique at 4.8~K ($A_{\rm mod}$ = 15 meV, $f$ = 531 Hz). \\

\subsection*{FERT spectroscopy}
Field emission resonance tunneling d$Z$/d$V$ were obtained with lock-in modulation technique where the modulation frequency and amplitude were equal to $f_0$ = 610 Hz and $A_{\rm mod}$= 50 mV, respectively. FERT spectra were measured with $Z$-feedback on, thus the tip-sample distance varied to keep constant the current \cite{Chahib2024}.

\subsection*{Dissipation spectroscopy}
 Frequency shift $\Delta f(V)$ and dissipation $\Gamma(V)$ has been recorded in dynamic AFM mode, by oscillating tuning fork at the oscillation amplitude $A$ = 100 pm. The cantilever damping was calculated according to the formula:

\begin{equation}
    \Gamma(\textrm{kg/s})=\Gamma_0(\frac{A_{\textrm{exc}}(z)}{A_{\textrm{exc,0}}}-\frac{f(z)}{f_0})
\end{equation}

where $A_{exc}(z)$ and $f(z)$ are the distance-dependent excitation amplitude and resonance frequency of the cantilever, and the suffix 0 refers to the free cantilever. The distance $Z$ = 0 corresponds to the point where the tip enters the contact regime, meaning that the cantilever driving signal is saturated and the tunneling current starts to rise. $\Gamma_0$ is the free cantilever internal dissipation given by: 

\begin{equation}
    \Gamma_0(\textrm{kg/s})=\frac{k}{2\pi f_0 Q}
\end{equation}

where $Q$ = 2000 is the quality factor of the sensor.
The damping can be converted into energy dissipation using the conversion formula:

\begin{equation}
    P(\textrm{eV/cycle})=\frac{2 \pi^2A^2f_0}{e}~\Gamma(\textrm{kg/s})
\end{equation}

\subsection*{pAFM spectroscopy}
The B-field dependent experiments have been performed in a separate low temperature pendulum Atomic Force Microscopy (pAFM) system. Due to its low spring constant $k$ = 0.1 Nm$^{-1}$ and high $Q$ = 7 $\times$ 10$^{5}$, the force sensor in the pAFM geometry offers enhanced force and dissipation sensitivity. We used highly n-doped silicon cantilever (ATEC-Cont from Nanosensors with resistivity $\rho$ = 0.01-0.02 $\Omega$cm). The pAFM sensor is suspended perpendicularly to the surface, meaning that the tip vibrational motion is parallel to the sample surface. The oscillation amplitude $A$  was kept constant by means of a phase-locked loop feedback system at 5 nm. Cantilevers were annealed in UHV up to 700 °C for 12h to remove surface contaminants from both cantilever and tip. It is also known that this long-term annealing leads to negligible amounts of localized charges on the probing tip. To compensate the contact potential difference between tip and sample, the voltage was applied to the sample. The magnetic field was applied out of surface plane, along the pAFM cantilever axis.

\section*{Declarations}
\backmatter
\bmhead{Supplementary information}

The Supporting Information is available free of charge as pdf file.

\bmhead{Acknowledgments}
We gratefully acknowledge the Werner Siemens Stiftung (WSS) for supporting the WSS Research Centre for Molecular Quantum Systems (molQ). A.G., E.M., M.K. and R.P. acknowledge funding from the Swiss Nanoscience Institute (SNI) and the European Research Council (ERC) under the European Union’s Horizon 2020 research and innovation programme (ULTRADISS) grant agreement No 834402 and supports as a part of NCCR SPIN, a National Centre of Competence (or Excellence) in Research. E.M., M.K. and R.P. acknowledge the SNF grant (200020\_228403).

\bmhead{Funding}
The work was funded by National Center of Competence (or Excellence) in Research under the SNF (grant number 51NF40-180604) and the SNF grant (200020\_228403).

\bmhead{Data availability}
The data underlying this article are available in Zenodo at 10.5281/zenodo.17454248

\bmhead{Author contributions}
M.K, R.P. and E.M. planned the experiments.
A.G., M.K and R.P. performed the experiments.
A.G. and M.K analyzed the data.
A.G., M.K and R.P. wrote the manuscript.
All authors discussed on the results and revised the manuscript.

\bmhead{Competing interests}
The authors declare no competing interests.


\begin{thebibliography}{99}
\expandafter\ifx\csname url\endcsname\relax
  \def\url#1{\burl{#1}}\fi
\expandafter\ifx\csname urlprefix\endcsname\relax\def\urlprefix{URL }\fi
\providecommand{\bibinfo}[2]{#2}
\providecommand{\eprint}[2][]{\url{#2}}
\providecommand{\doi}[1]{\url{https://doi.org/#1}}
\bibcommenthead

\bibitem{Mueller1979}
\bibinfo{author}{Müller, K.~A.} \& \bibinfo{author}{Burkard, H.}
\newblock \bibinfo{title}{Srti${\mathrm{o}}_{3}$: An intrinsic quantum
  paraelectric below 4 k}.
\newblock \emph{\bibinfo{journal}{Phys. Rev. B}}
  \textbf{\bibinfo{volume}{19}}~(7), \bibinfo{pages}{3593--3602}
  (\bibinfo{year}{1979}).
\newblock \doi{10.1103/PhysRevB.19.3593} .

\bibitem{Uwe1976}
\bibinfo{author}{Uwe, H.} \& \bibinfo{author}{Sakudo, T.}
\newblock \bibinfo{title}{Stress-induced ferroelectricity and soft phonon modes
  in srti${\mathrm{o}}_{3}$}.
\newblock \emph{\bibinfo{journal}{Phys. Rev. B}}
  \textbf{\bibinfo{volume}{13}}~(1), \bibinfo{pages}{271--286}
  (\bibinfo{year}{1976}).
\newblock \doi{10.1103/PhysRevB.13.271} .

\bibitem{Zhang2015}
\bibinfo{author}{Zhang, Y.}, \bibinfo{author}{Wang, J.},
  \bibinfo{author}{Sahoo, M.}, \bibinfo{author}{Shimada, T.} \&
  \bibinfo{author}{Kitamura, T.}
\newblock \bibinfo{title}{Mechanical control of magnetism in oxygen deficient
  perovskite {SrTiO}$_{\textrm{3}}$}.
\newblock \emph{\bibinfo{journal}{Phys. Chem. Chem. Phys.}}
  \textbf{\bibinfo{volume}{17}}~(40), \bibinfo{pages}{27136--27144}
  (\bibinfo{year}{2015}).
\newblock \doi{10.1039/C5CP04310G} .

\bibitem{Fleury1968}
\bibinfo{author}{Fleury, P.~A.} \& \bibinfo{author}{Worlock, J.~M.}
\newblock \bibinfo{title}{Electric-field-induced raman scattering in
  srti${\mathrm{o}}_{3}$ and kta${\mathrm{o}}_{3}$}.
\newblock \emph{\bibinfo{journal}{Phys. Rev.}}
  \textbf{\bibinfo{volume}{174}}~(2), \bibinfo{pages}{613--623}
  (\bibinfo{year}{1968}).
\newblock \doi{10.1103/PhysRev.174.613} .

\bibitem{Bickel1989}
\bibinfo{author}{Bickel, N.}, \bibinfo{author}{Schmidt, G.},
  \bibinfo{author}{Heinz, K.} \& \bibinfo{author}{Müller, K.}
\newblock \bibinfo{title}{Ferroelectric relaxation of the
  ${\mathrm{srtio}}_{3}$(100) surface}.
\newblock \emph{\bibinfo{journal}{Phys. Rev. Lett.}}
  \textbf{\bibinfo{volume}{62}}~(17), \bibinfo{pages}{2009--2011}
  (\bibinfo{year}{1989}).
\newblock \doi{10.1103/PhysRevLett.62.2009} .

\bibitem{Santander2011}
\bibinfo{author}{Santander-Syro, A.~F.} \emph{et~al.}
\newblock \bibinfo{title}{Two-dimensional electron gas with universal subbands
  at the surface of {SrTiO3}}.
\newblock \emph{\bibinfo{journal}{Nature}}
  \textbf{\bibinfo{volume}{469}}~(7329), \bibinfo{pages}{189--193}
  (\bibinfo{year}{2011}).
\newblock \doi{10.1038/nature09720} .

\bibitem{Santander2014}
\bibinfo{author}{Santander-Syro, A.~F.} \emph{et~al.}
\newblock \bibinfo{title}{Giant spin splitting of the two-dimensional electron
  gas at the surface of {SrTiO3}}.
\newblock \emph{\bibinfo{journal}{Nat. Mater.}}
  \textbf{\bibinfo{volume}{13}}~(12), \bibinfo{pages}{1085--1090}
  (\bibinfo{year}{2014}).
\newblock \doi{10.1038/nmat4107} .

\bibitem{Sokolovic2025}
\bibinfo{author}{Sokolović, I.} \emph{et~al.}
\newblock \bibinfo{title}{Duality and degeneracy lifting in two-dimensional
  electron liquids on srtio3(001)}.
\newblock \emph{\bibinfo{journal}{Nat. Commun.}}
  \textbf{\bibinfo{volume}{16}}~(1), \bibinfo{pages}{4594}
  (\bibinfo{year}{2025}).
\newblock \doi{10.1038/s41467-025-59258-4} .

\bibitem{Meevasana2011}
\bibinfo{author}{Meevasana, W.} \emph{et~al.}
\newblock \bibinfo{title}{Creation and control of a two-dimensional electron
  liquid at the bare srtio3 surface}.
\newblock \emph{\bibinfo{journal}{Nat. Mater.}}
  \textbf{\bibinfo{volume}{10}}~(2), \bibinfo{pages}{114--118}
  (\bibinfo{year}{2011}).
\newblock \doi{10.1038/nmat2943} .

\bibitem{Plumb2014}
\bibinfo{author}{Plumb, N.~C.} \emph{et~al.}
\newblock \bibinfo{title}{Mixed dimensionality of confined conducting electrons
  in the surface region of ${\mathrm{srtio}}_{3}$}.
\newblock \emph{\bibinfo{journal}{Phys. Rev. Lett.}}
  \textbf{\bibinfo{volume}{113}}, \bibinfo{pages}{086801}
  (\bibinfo{year}{2014}).
\newblock \doi{10.1103/PhysRevLett.113.086801} .

\bibitem{Ohta2007}
\bibinfo{author}{Ohta, H.} \emph{et~al.}
\newblock \bibinfo{title}{Giant thermoelectric {Seebeck} coefficient of a
  two-dimensional electron gas in {SrTiO3}}.
\newblock \emph{\bibinfo{journal}{Nat. Mater.}}
  \textbf{\bibinfo{volume}{6}}~(2), \bibinfo{pages}{129--134}
  (\bibinfo{year}{2007}).
\newblock \doi{10.1038/nmat1821} .

\bibitem{Annadi2013}
\bibinfo{author}{Annadi, A.} \emph{et~al.}
\newblock \bibinfo{title}{Anisotropic two-dimensional electron gas at the
  {LaAlO3}/{SrTiO3} (110) interface}.
\newblock \emph{\bibinfo{journal}{Nat. Commun.}}
  \textbf{\bibinfo{volume}{4}}~(1), \bibinfo{pages}{1838}
  (\bibinfo{year}{2013}).
\newblock \doi{10.1038/ncomms2804} .

\bibitem{Ohtomo2004}
\bibinfo{author}{Ohtomo, A.} \& \bibinfo{author}{Hwang, H.~Y.}
\newblock \bibinfo{title}{A high-mobility electron gas at the {LaAlO3}/{SrTiO3}
  heterointerface}.
\newblock \emph{\bibinfo{journal}{Nature}}
  \textbf{\bibinfo{volume}{427}}~(6973), \bibinfo{pages}{423--426}
  (\bibinfo{year}{2004}).
\newblock \doi{10.1038/nature02308} .

\bibitem{Taniuchi2016}
\bibinfo{author}{Taniuchi, T.} \emph{et~al.}
\newblock \bibinfo{title}{Imaging of room-temperature ferromagnetic
  nano-domains at the surface of a non-magnetic oxide}.
\newblock \emph{\bibinfo{journal}{Nat. Commun.}}
  \textbf{\bibinfo{volume}{7}}~(1), \bibinfo{pages}{11781}
  (\bibinfo{year}{2016}).
\newblock \doi{10.1038/ncomms11781} .

\bibitem{Coey2016}
\bibinfo{author}{Coey, J. M.~D.}, \bibinfo{author}{Venkatesan, M.} \&
  \bibinfo{author}{Stamenov, P.}
\newblock \bibinfo{title}{Surface magnetism of strontium titanate}.
\newblock \emph{\bibinfo{journal}{J. Phys. Cond. Matt.}}
  \textbf{\bibinfo{volume}{28}}~(48), \bibinfo{pages}{485001}
  (\bibinfo{year}{2016}).
\newblock \urlprefix\url{https://dx.doi.org/10.1088/0953-8984/28/48/485001}.
\newblock \doi{10.1088/0953-8984/28/48/485001} .

\bibitem{Pavlenko2013}
\bibinfo{author}{Pavlenko, N.} \& \bibinfo{author}{Kopp, T.}
\newblock \bibinfo{title}{Magnetically {Ordered} {State} at {Correlated}
  {Oxide} {Interfaces}: {The} {Role} of {Random} {Oxygen} {Defects}}.
\newblock \emph{\bibinfo{journal}{Journal of Superconductivity and Novel
  Magnetism}} \textbf{\bibinfo{volume}{26}}~(4), \bibinfo{pages}{1175--1178}
  (\bibinfo{year}{2013}).
\newblock \doi{10.1007/s10948-012-2045-8} .

\bibitem{Schooley1964}
\bibinfo{author}{Schooley, J.~F.}, \bibinfo{author}{Hosler, W.~R.} \&
  \bibinfo{author}{Cohen, M.~L.}
\newblock \bibinfo{title}{Superconductivity in {Semiconducting} {SrTi} {O} 3}.
\newblock \emph{\bibinfo{journal}{Phys. Rev. Lett.}}
  \textbf{\bibinfo{volume}{12}}~(17), \bibinfo{pages}{474--475}
  (\bibinfo{year}{1964}).
\newblock \doi{10.1103/PhysRevLett.12.474} .

\bibitem{King2014}
\bibinfo{author}{King, P. D.~C.} \emph{et~al.}
\newblock \bibinfo{title}{Quasiparticle dynamics and spin–orbital texture of
  the {SrTiO3} two-dimensional electron gas}.
\newblock \emph{\bibinfo{journal}{Nat. Commun.}}
  \textbf{\bibinfo{volume}{5}}~(1), \bibinfo{pages}{3414}
  (\bibinfo{year}{2014}).
\newblock \doi{10.1038/ncomms4414} .

\bibitem{Caviglia2010}
\bibinfo{author}{Caviglia, A.~D.} \emph{et~al.}
\newblock \bibinfo{title}{Tunable rashba spin-orbit interaction at oxide
  interfaces}.
\newblock \emph{\bibinfo{journal}{Phys. Rev. Lett.}}
  \textbf{\bibinfo{volume}{104}}~(12), \bibinfo{pages}{126803}
  (\bibinfo{year}{2010}).
\newblock \doi{10.1103/PhysRevLett.104.126803} .

\bibitem{Fete2014}
\bibinfo{author}{Fête, A.} \emph{et~al.}
\newblock \bibinfo{title}{Large modulation of the shubnikov-de haas
  oscillations by the rashba interaction at the laalo3/srtio3 interface}.
\newblock \emph{\bibinfo{journal}{New Jour. Phys.}}
  \textbf{\bibinfo{volume}{16}}~(11), \bibinfo{pages}{112002}
  (\bibinfo{year}{2014}).
\newblock \doi{10.1088/1367-2630/16/11/112002} .

\bibitem{Li2019}
\bibinfo{author}{Li, X.} \emph{et~al.}
\newblock \bibinfo{title}{Terahertz field-induced ferroelectricity in quantum
  paraelectric srtio3}.
\newblock \emph{\bibinfo{journal}{Science}}
  \textbf{\bibinfo{volume}{364}}~(6445), \bibinfo{pages}{1079--1082}
  (\bibinfo{year}{2019}).
\newblock \doi{10.1126/science.aaw4913} .

\bibitem{Cen2008}
\bibinfo{author}{Cen, C.} \emph{et~al.}
\newblock \bibinfo{title}{Nanoscale control of an interfacial metal–insulator
  transition at room temperature}.
\newblock \emph{\bibinfo{journal}{Nat. Mater.}}
  \textbf{\bibinfo{volume}{7}}~(4), \bibinfo{pages}{298--302}
  (\bibinfo{year}{2008}).
\newblock \doi{10.1038/nmat2136} .

\bibitem{Cen2009}
\bibinfo{author}{Cen, C.}, \bibinfo{author}{Thiel, S.},
  \bibinfo{author}{Mannhart, J.} \& \bibinfo{author}{Levy, J.}
\newblock \bibinfo{title}{Oxide {Nanoelectronics} on {Demand}}.
\newblock \emph{\bibinfo{journal}{Science}}
  \textbf{\bibinfo{volume}{323}}~(5917), \bibinfo{pages}{1026--1030}
  (\bibinfo{year}{2009}).
\newblock \doi{10.1126/science.1168294} .

\bibitem{Gerhold2014}
\bibinfo{author}{Gerhold, S.}, \bibinfo{author}{Wang, Z.},
  \bibinfo{author}{Schmid, M.} \& \bibinfo{author}{Diebold, U.}
\newblock \bibinfo{title}{Stoichiometry-driven switching between surface
  reconstructions on srtio3(001)}.
\newblock \emph{\bibinfo{journal}{Surf. Sci.}} \textbf{\bibinfo{volume}{621}},
  \bibinfo{pages}{L1--L4} (\bibinfo{year}{2014}).
\newblock \doi{10.1016/j.susc.2013.10.015} .

\bibitem{Dagdeviren2016}
\bibinfo{author}{Dagdeviren, O.~E.} \emph{et~al.}
\newblock \bibinfo{title}{Surface phase, morphology, and charge distribution
  transitions on vacuum and ambient annealed
  \${\textbackslash}mathrm\{{SrTi}\}\{{\textbackslash}mathrm\{{O}\}\}\_\{3\}\$(100)}.
\newblock \emph{\bibinfo{journal}{Phys. Rev. B}}
  \textbf{\bibinfo{volume}{93}}~(19), \bibinfo{pages}{195303}
  (\bibinfo{year}{2016}).
\newblock \doi{10.1103/PhysRevB.93.195303}, \bibinfo{note}{publisher: American
  Physical Society} .

\bibitem{Tanaka1993}
\bibinfo{author}{Tanaka, H.}, \bibinfo{author}{Matsumoto, T.},
  \bibinfo{author}{Tomoji~Kawai, T.~K.} \& \bibinfo{author}{Shichio~Kawai,
  S.~K.}
\newblock \bibinfo{title}{Surface {Structure} and {Electronic} {Property} of
  {Reduced} {SrTiO}$_{\textrm{3}}$ (100) {Surface} {Observed} by {Scanning}
  {Tunneling} {Microscopy}/{Spectroscopy}}.
\newblock \emph{\bibinfo{journal}{Japan. Jour. Appl. Phys.}}
  \textbf{\bibinfo{volume}{32}}~(3S), \bibinfo{pages}{1405}
  (\bibinfo{year}{1993}).
\newblock \doi{10.1143/JJAP.32.1405} .

\bibitem{Kisiel2018a}
\bibinfo{author}{Kisiel, M.} \emph{et~al.}
\newblock \bibinfo{title}{Mechanical dissipation from charge and spin
  transitions in oxygen-deficient {SrTiO3} surfaces}.
\newblock \emph{\bibinfo{journal}{Nat. Commun.}}
  \textbf{\bibinfo{volume}{9}}~(1), \bibinfo{pages}{2946}
  (\bibinfo{year}{2018}).
\newblock \doi{10.1038/s41467-018-05392-1} .

\bibitem{Brovko2017}
\bibinfo{author}{Brovko, O.~O.} \& \bibinfo{author}{Tosatti, E.}
\newblock \bibinfo{title}{Controlling the magnetism of oxygen surface vacancies
  in ${\mathrm{srtio}}_{3}$ through charging}.
\newblock \emph{\bibinfo{journal}{Phys. Rev. Mater.}}
  \textbf{\bibinfo{volume}{1}}, \bibinfo{pages}{044405} (\bibinfo{year}{2017}).
\newblock \doi{10.1103/PhysRevMaterials.1.044405} .

\bibitem{DiCapua2012}
\bibinfo{author}{Di~Capua, R.} \emph{et~al.}
\newblock \bibinfo{title}{Observation of a two-dimensional electron gas at the
  surface of annealed srtio${}_{3}$ single crystals by scanning tunneling
  spectroscopy}.
\newblock \emph{\bibinfo{journal}{Phys. Rev. B}} \textbf{\bibinfo{volume}{86}},
  \bibinfo{pages}{155425} (\bibinfo{year}{2012}).
\newblock \doi{10.1103/PhysRevB.86.155425} .

\bibitem{Guedes2021}
\bibinfo{author}{Guedes, E.~B.} \emph{et~al.}
\newblock \bibinfo{title}{Universal {Structural} {Influence} on the {2D}
  {Electron} {Gas} at {SrTiO3} {Surfaces}}.
\newblock \emph{\bibinfo{journal}{Adv. Sci.}}
  \textbf{\bibinfo{volume}{8}}~(22), \bibinfo{pages}{2100602}
  (\bibinfo{year}{2021}).
\newblock \doi{10.1002/advs.202100602}, \bibinfo{note}{\_eprint:
  https://onlinelibrary.wiley.com/doi/pdf/10.1002/advs.202100602} .

\bibitem{Altmeyer2016}
\bibinfo{author}{Altmeyer, M.} \emph{et~al.}
\newblock \bibinfo{title}{Magnetism, spin texture, and in-gap states: Atomic
  specialization at the surface of oxygen-deficient ${\mathrm{srtio}}_{3}$}.
\newblock \emph{\bibinfo{journal}{Phys. Rev. Lett.}}
  \textbf{\bibinfo{volume}{116}}, \bibinfo{pages}{157203}
  (\bibinfo{year}{2016}).
\newblock \doi{10.1103/PhysRevLett.116.157203} .

\bibitem{Huetner2024}
\bibinfo{author}{Hütner, J.~I.} \emph{et~al.}
\newblock \bibinfo{title}{Stoichiometric reconstruction of the al2o3(0001)
  surface}.
\newblock \emph{\bibinfo{journal}{Science}}
  \textbf{\bibinfo{volume}{385}}~(6714), \bibinfo{pages}{1241--1244}
  (\bibinfo{year}{2024}).
\newblock \doi{10.1126/science.adq4744} .

\bibitem{Sokolovic2019}
\bibinfo{author}{Sokolović, I.}, \bibinfo{author}{Schmid, M.},
  \bibinfo{author}{Diebold, U.} \& \bibinfo{author}{Setvin, M.}
\newblock \bibinfo{title}{Incipient ferroelectricity: A route towards
  bulk-terminated ${\mathrm{srtio}}_{3}$}.
\newblock \emph{\bibinfo{journal}{Phys. Rev. Mater.}}
  \textbf{\bibinfo{volume}{3}}~(3), \bibinfo{pages}{034407}
  (\bibinfo{year}{2019}).
\newblock
  \urlprefix\url{https://link.aps.org/doi/10.1103/PhysRevMaterials.3.034407}.
\newblock \doi{10.1103/PhysRevMaterials.3.034407} .

\bibitem{Kisiel2015}
\bibinfo{author}{Kisiel, M.} \emph{et~al.}
\newblock \bibinfo{title}{Noncontact dissipation reveals critical central peak
  in {SrTiO3}}.
\newblock \emph{\bibinfo{journal}{Phys. Rev. Lett.}}
  \textbf{\bibinfo{volume}{115}}~(4), \bibinfo{pages}{046101}
  (\bibinfo{year}{2015}).
\newblock \doi{10.1103/PhysRevLett.115.046101}, \bibinfo{note}{arXiv:1506.01306
  [cond-mat]} .

\bibitem{Stomp2005}
\bibinfo{author}{Stomp, R.} \emph{et~al.}
\newblock \bibinfo{title}{Detection of {Single}-{Electron} {Charging} in an
  {Individual} {InAs} {Quantum} {Dot} by {Noncontact} {Atomic}-{Force}
  {Microscopy}}.
\newblock \emph{\bibinfo{journal}{Phys. Rev. Lett.}}
  \textbf{\bibinfo{volume}{94}}~(5), \bibinfo{pages}{056802}
  (\bibinfo{year}{2005}).
\newblock \doi{10.1103/PhysRevLett.94.056802} .

\bibitem{Gross2009}
\bibinfo{author}{Gross, L.} \emph{et~al.}
\newblock \bibinfo{title}{Measuring the {Charge} {State} of an {Adatom} with
  {Noncontact} {Atomic} {Force} {Microscopy}}.
\newblock \emph{\bibinfo{journal}{Science}}
  \textbf{\bibinfo{volume}{324}}~(5933), \bibinfo{pages}{1428--1431}
  (\bibinfo{year}{2009}).
\newblock \doi{10.1126/science.1172273} .

\bibitem{Setvin2017}
\bibinfo{author}{Setvin, M.}, \bibinfo{author}{Hulva, J.},
  \bibinfo{author}{Parkinson, G.~S.}, \bibinfo{author}{Schmid, M.} \&
  \bibinfo{author}{Diebold, U.}
\newblock \bibinfo{title}{Electron transfer between anatase tio2 and an o2
  molecule directly observed by atomic force microscopy}.
\newblock \emph{\bibinfo{journal}{Proc. Nat. Acad. Sci.}}
  \textbf{\bibinfo{volume}{114}}~(13), \bibinfo{pages}{E2556--E2562}
  (\bibinfo{year}{2017}).
\newblock \doi{10.1073/pnas.1618723114} .

\bibitem{Kocic2015}
\bibinfo{author}{Kocić, N.} \emph{et~al.}
\newblock \bibinfo{title}{Periodic {Charging} of {Individual} {Molecules}
  {Coupled} to the {Motion} of an {Atomic} {Force} {Microscopy} {Tip}}.
\newblock \emph{\bibinfo{journal}{Nano Lett.}}
  \textbf{\bibinfo{volume}{15}}~(7), \bibinfo{pages}{4406--4411}
  (\bibinfo{year}{2015}).
\newblock \doi{10.1021/acs.nanolett.5b00711} .

\bibitem{Fatayer2019}
\bibinfo{author}{Fatayer, S.} \emph{et~al.}
\newblock \bibinfo{title}{Molecular structure elucidation with charge-state
  control}.
\newblock \emph{\bibinfo{journal}{Science}}
  \textbf{\bibinfo{volume}{365}}~(6449), \bibinfo{pages}{142--145}
  (\bibinfo{year}{2019}).
\newblock \urlprefix\url{https://doi.org/10.1126/science.aax5895}.
\newblock \doi{10.1126/science.aax5895} .

\bibitem{Li2023}
\bibinfo{author}{Li, C.} \emph{et~al.}
\newblock \bibinfo{title}{Strong signature of electron-vibration coupling in
  molecules on ag(111) triggered by tip-gated discharging}.
\newblock \emph{\bibinfo{journal}{Nat. Comm.}}
  \textbf{\bibinfo{volume}{14}}~(1), \bibinfo{pages}{5956}
  (\bibinfo{year}{2023}).
\newblock \doi{10.1038/s41467-023-41601-2} .

\bibitem{Scheuerer2020}
\bibinfo{author}{Scheuerer, P.}, \bibinfo{author}{Patera, L.~L.} \&
  \bibinfo{author}{Repp, J.}
\newblock \bibinfo{title}{Manipulating and probing the distribution of excess
  electrons in an electrically isolated self-assembled molecular structure}.
\newblock \emph{\bibinfo{journal}{Nano Lett.}}
  \textbf{\bibinfo{volume}{20}}~(3), \bibinfo{pages}{1839--1845}
  (\bibinfo{year}{2020}).
\newblock \urlprefix\url{https://doi.org/10.1021/acs.nanolett.9b05063}.
\newblock \doi{10.1021/acs.nanolett.9b05063} .

\bibitem{Dastolfo2022}
\bibinfo{author}{D’Astolfo, P.} \emph{et~al.}
\newblock \bibinfo{title}{Energy {Dissipation} from {Confined} {States} in
  {Nanoporous} {Molecular} {Networks}}.
\newblock \emph{\bibinfo{journal}{ACS Nano}}
  \textbf{\bibinfo{volume}{16}}~(10), \bibinfo{pages}{16314--16321}
  (\bibinfo{year}{2022}).
\newblock \doi{10.1021/acsnano.2c05333}, \bibinfo{note}{publisher: American
  Chemical Society} .

\bibitem{Ollier2023}
\bibinfo{author}{Ollier, A.} \emph{et~al.}
\newblock \bibinfo{title}{Energy dissipation on magic angle twisted bilayer
  graphene}.
\newblock \emph{\bibinfo{journal}{Comm. Phys.}}
  \textbf{\bibinfo{volume}{6}}~(1), \bibinfo{pages}{344}
  (\bibinfo{year}{2023}).
\newblock \doi{10.1038/s42005-023-01441-4} .

\bibitem{Dose1987}
\bibinfo{author}{Dose, V.}
\newblock \bibinfo{title}{Image potential surface states}.
\newblock \emph{\bibinfo{journal}{Phys. Scrip.}}
  \textbf{\bibinfo{volume}{36}}~(4), \bibinfo{pages}{669--672}
  (\bibinfo{year}{1987}).
\newblock \doi{10.1088/0031-8949/36/4/009} .

\bibitem{Wahl2003}
\bibinfo{author}{Wahl, P.}, \bibinfo{author}{Schneider, M.~A.},
  \bibinfo{author}{Diekhöner, L.}, \bibinfo{author}{Vogelgesang, R.} \&
  \bibinfo{author}{Kern, K.}
\newblock \bibinfo{title}{Quantum {Coherence} of {Image}-{Potential} {States}}.
\newblock \emph{\bibinfo{journal}{Phys. Rev. Lett.}}
  \textbf{\bibinfo{volume}{91}}~(10), \bibinfo{pages}{106802}
  (\bibinfo{year}{2003}).
\newblock \doi{10.1103/PhysRevLett.91.106802} .

\bibitem{Yildiz2019}
\bibinfo{author}{Yildiz, D.}, \bibinfo{author}{Kisiel, M.},
  \bibinfo{author}{Gysin, U.}, \bibinfo{author}{Gürlü, O.} \&
  \bibinfo{author}{Meyer, E.}
\newblock \bibinfo{title}{Mechanical dissipation via image potential states on
  a topological insulator surface}.
\newblock \emph{\bibinfo{journal}{Nat. Mater.}}
  \textbf{\bibinfo{volume}{18}}~(11), \bibinfo{pages}{1201--1206}
  (\bibinfo{year}{2019}).
\newblock \doi{10.1038/s41563-019-0492-3} .

\bibitem{Chahib2024}
\bibinfo{author}{Chahib, O.} \emph{et~al.}
\newblock \bibinfo{title}{Probing charge redistribution at the interface of
  self-assembled cyclo-{P5} pentamers on {Ag}(111)}.
\newblock \emph{\bibinfo{journal}{Nat. Commun.}}
  \textbf{\bibinfo{volume}{15}}~(1), \bibinfo{pages}{6542}
  (\bibinfo{year}{2024}).
\newblock \doi{10.1038/s41467-024-50862-4} .

\bibitem{he_impurity_2011}
\bibinfo{author}{He, H.-T.} \emph{et~al.}
\newblock \bibinfo{title}{Impurity {Effect} on {Weak} {Antilocalization} in the
  {Topological} {Insulator} {Bi} 2 {Te} 3}.
\newblock \emph{\bibinfo{journal}{Phys. Rev. Lett..}}
  \textbf{\bibinfo{volume}{106}}~(16), \bibinfo{pages}{166805}
  (\bibinfo{year}{2011}).
\newblock \doi{10.1103/PhysRevLett.106.166805} .

\bibitem{takeyasu_control_2013}
\bibinfo{author}{Takeyasu, K.}, \bibinfo{author}{Fukada, K.},
  \bibinfo{author}{Matsumoto, M.} \& \bibinfo{author}{Fukutani, K.}
\newblock \bibinfo{title}{Control of the surface electronic structure of
  {SrTiO}$_{\textrm{3}}$ (001) by modulation of the density of oxygen
  vacancies}.
\newblock \emph{\bibinfo{journal}{J. Phys.: Cond. Matter}}
  \textbf{\bibinfo{volume}{25}}~(16), \bibinfo{pages}{162202}
  (\bibinfo{year}{2013}).
\newblock \doi{10.1088/0953-8984/25/16/162202} .

\bibitem{Chung1979}
\bibinfo{author}{Chung, Y.-W.} \& \bibinfo{author}{Weissbard, W.~B.}
\newblock \bibinfo{title}{Surface spectroscopy studies of the
  srti${\mathrm{o}}_{3}$ (100) surface and the platinum-srti${\mathrm{o}}_{3}$
  (100) interface}.
\newblock \emph{\bibinfo{journal}{Phys. Rev. B}}
  \textbf{\bibinfo{volume}{20}}~(8), \bibinfo{pages}{3456--3461}
  (\bibinfo{year}{1979}).
\newblock \urlprefix\url{https://link.aps.org/doi/10.1103/PhysRevB.20.3456}.
\newblock \doi{10.1103/PhysRevB.20.3456} .

\bibitem{Luryi1988}
\bibinfo{author}{Luryi, S.}
\newblock \bibinfo{title}{Quantum capacitance devices}.
\newblock \emph{\bibinfo{journal}{Appl. Phys. Lett.}}
  \textbf{\bibinfo{volume}{52}}~(6), \bibinfo{pages}{501--503}
  (\bibinfo{year}{1988}).
\newblock \doi{10.1063/1.99649} .

\bibitem{Kisiel2011}
\bibinfo{author}{Kisiel, M.} \emph{et~al.}
\newblock \bibinfo{title}{Suppression of electronic friction on {Nb} films in
  the superconducting state}.
\newblock \emph{\bibinfo{journal}{Nat. Mater.}}
  \textbf{\bibinfo{volume}{10}}~(2), \bibinfo{pages}{119--122}
  (\bibinfo{year}{2011}).
\newblock \doi{10.1038/nmat2936} .

\bibitem{Sader2004}
\bibinfo{author}{Sader, J.~E.} \& \bibinfo{author}{Jarvis, S.~P.}
\newblock \bibinfo{title}{Accurate formulas for interaction force and energy in
  frequency modulation force spectroscopy}.
\newblock \emph{\bibinfo{journal}{Appl. Phys. Lett.}}
  \textbf{\bibinfo{volume}{84}}~(10), \bibinfo{pages}{1801--1803}
  (\bibinfo{year}{2004}).
\newblock \doi{10.1063/1.1667267} .

\bibitem{volokitin_giant_2007}
\bibinfo{author}{Volokitin, A.~I.}, \bibinfo{author}{Persson, B. N.~J.} \&
  \bibinfo{author}{Ueba, H.}
\newblock \bibinfo{title}{Giant enhancement of noncontact friction between
  closely spaced bodies by dielectric films and two-dimensional systems}.
\newblock \emph{\bibinfo{journal}{Journal of Experimental and Theoretical
  Physics}} \textbf{\bibinfo{volume}{104}}~(1), \bibinfo{pages}{96--110}
  (\bibinfo{year}{2007}).
\newblock \urlprefix\url{http://link.springer.com/10.1134/S1063776107010116}.
\newblock \doi{10.1134/S1063776107010116} .

\bibitem{kim_quantum_2022}
\bibinfo{author}{Kim, S.} \emph{et~al.}
\newblock \bibinfo{title}{Quantum electron liquid and its possible phase
  transition}.
\newblock \emph{\bibinfo{journal}{Nature Materials}}
  \textbf{\bibinfo{volume}{21}}~(11), \bibinfo{pages}{1269--1274}
  (\bibinfo{year}{2022}).
\newblock \urlprefix\url{https://www.nature.com/articles/s41563-022-01353-8}.
\newblock \doi{10.1038/s41563-022-01353-8} .

\bibitem{kingQuasiparticleDynamicsSpin2014a}
\bibinfo{author}{King, P. D.~C.} \emph{et~al.}
\newblock \bibinfo{title}{Quasiparticle dynamics and spin–orbital texture of
  the {SrTiO3} two-dimensional electron gas}.
\newblock \emph{\bibinfo{journal}{Nat. Commun.}}
  \textbf{\bibinfo{volume}{5}}~(1), \bibinfo{pages}{3414}
  (\bibinfo{year}{2014}).
\newblock \doi{10.1038/ncomms4414} .

\bibitem{sanden_quantum_1987}
\bibinfo{author}{Sanden, M. C. M. V.~D.}, \bibinfo{author}{Heijden, R. W.
  V.~D.}, \bibinfo{author}{Waele, A. T. A. M.~D.} \& \bibinfo{author}{Gijsman,
  H.~M.}
\newblock \bibinfo{title}{Quantum {Magnetoconductance} of the
  {Two}-{Dimensional} {Electron} {Gas} on a {Liquid} {Helium} {Surface}}.
\newblock \emph{\bibinfo{journal}{Japanese Journal of Applied Physics}}
  \textbf{\bibinfo{volume}{26}}~(S3-1), \bibinfo{pages}{749}
  (\bibinfo{year}{1987}).
\newblock
  \urlprefix\url{https://iopscience.iop.org/article/10.7567/JJAPS.26S3.749}.
\newblock \doi{10.7567/JJAPS.26S3.749} .

\bibitem{Giessibl2019}
\bibinfo{author}{Giessibl, F.~J.}
\newblock \bibinfo{title}{The {qPlus} sensor, a powerful core for the atomic
  force microscope}.
\newblock \emph{\bibinfo{journal}{Rev. Sci. Instr.}}
  \textbf{\bibinfo{volume}{90}}~(1), \bibinfo{pages}{011101}
  (\bibinfo{year}{2019}).
\newblock \doi{10.1063/1.5052264} .

\end{thebibliography}

\end{document}